\documentclass[preprint,aps,preprintnumbers,nofootinbib,showpacs]{revtex4}

\begin{document}

\title{A Simple Model for Quintessential Inflation}

\author{R. Rosenfeld}
\email{rosenfel@ift.unesp.br}
\affiliation{ 
Instituto de F\'{\i}sica Te\'orica - UNESP, Rua Pamplona, 145, 01405-900, 
S\~{a}o Paulo, SP, Brazil} 
\author{J. A. Frieman} 
\email{frieman@fnal.gov}
\affiliation{NASA/Fermilab Astrophysics Center \\
Fermi National Accelerator Laboratory, Batavia, IL 60510 \\and\\
Department of Astronomy \& Astrophysics and Kavli Institute for
Cosmological Physics \\ The University of Chicago, Chicago, IL 60637}

\begin{abstract}
We propose a simple toy model for quintessential inflation where a complex scalar 
field described by a lagrangian with
a $U(1)_{PQ}$ symmetry spontaneously broken at a
high energy scale and explicitly broken by instanton effects at a much lower
energy can account for both the early inflationary phase and the recent
accelerated expansion of the Universe. The real part of the complex field plays
the role of the inflaton whereas the imaginary part, the ``axion", is the
quintessence field. 

\end{abstract}



\maketitle

\section{Introduction}

Recent observations of the cosmic microwave background \cite{wmap}, type Ia
supernovae \cite{snIa} and the large scale structure of the Universe \cite{sdss}
have provided evidence that the Universe has experienced two
different stages of accelerated expansion during its history. 
The inflationary stage took place at
the very first moments of the Universe and superluminally stretched a tiny patch
to become our observable Universe today. It generated the fluctuations 
responsible to seed the observed large scale structure and naturally explains
why the Universe is flat, homogeneous and isotropic. It ended with a period of
reheating that brought the Universe back to the usual Friedman-Robertson-Walker
expansion. The second stage of
accelerated expansion began at a redshift of order one-half and is still
operative. We do not know whether it will halt in the future. 

In order to physically generate an accelerated expansion 
in the framework of general relativity it is necessary to
invoke some form of vacuum energy, the most popular candidates being either 
a cosmological constant or a scalar field, sometimes called
quintessence, displaced from the minimum of its 
potential energy (which can be adjusted to vanish) \cite{ratra&peebles}.
Since inflation has to end, a cosmological constant cannot be its source.
Furthermore, it requires a tremendous amount of fine tuning in order to explain
the current acceleration stage with a cosmological constant.
Therefore, it is more natural to explore models of scalar fields in order to
explain the two stages of acceleration. 

In principle, these two stages of acceleration are independent. They occurred at
wildly different epochs of the Universe and may have been caused by totally 
unrelated physical processes. On the other hand, it would be most interesting 
if these 
two stages were in some way connected to each other. It is possible to develop 
the so-called quintessential inflation models, where a carefully crafted
single real scalar field potential can be responsible for both stages
\cite{QuintInf}. 
However, these models involve non-renormalizable, {\it ad hoc}
potentials. Furthermore, these potentials usually do not have a local minimum,
making the conventional reheating process unoperative. Particle production via
preheating is more efficient than gravitational particle production in this
case, but requires the introduction of extra fields \cite{us}.
In this paper we propose to use a simple, well motivated and renormalizable, 
complex scalar field model to unify the description of both stages of 
acceleration into a single framework. 

\section{The model}

In our proposal, we use a complex
scalar field pertaining to a model with a global $U(1)_{PQ}$-like 
symmetry spontaneously broken at a
high energy scale $f$ \cite{axions}.
This generates a flat potential for the imaginary part of the field (``axion")
that is lifted by small instanton effects at a much lower energy scale.
Essentially the same type of model has been proposed as a candidate for natural
inflation \cite{NatInf} and, separately, for quintessence \cite{Quint}, 
with different parameters in each case.
Also, complex scalar fields have been recently studied in the context of 
quintessence models called ``spintessence" \cite{spintessence}. 

We use a Lagrangian given by:
\begin{equation} \label{eq:lagrangian1}
{\cal L} =  \partial_\mu \Phi \partial^\mu \Phi^\ast - 
V(\Phi) - M^4 [ \cos(\mbox{Arg} (\Phi)) -1 ]
\end{equation}
with a renormalizable potential
\begin{equation}
V(\Phi) = \lambda \left( \Phi \Phi^\ast - \frac{f^2}{2}\right)^2.
\end{equation}

The global $U(1)_{PQ}$ symmetry is spontaneously broken at a high energy scale
determined by $f = {\cal O} (M_{Pl})$ and explicitly broken  
by small instanton effects at a much lower energy scale $M$.
The real and imaginary parts of the field $\Phi$ will
be identified with the inflaton and the quintessence fields, respectively. 

Writing the complex field $\Phi$ as
\begin{equation} 
\Phi = \frac{1}{\sqrt{2}} \eta e^{i \varphi/f}
\end{equation}
we note that, in order that the field $\varphi$ has not dissipated away its
energy density today producing axions, we should require:
\begin{equation}
m_{\varphi}  = \frac{M^2}{f} \lesssim 3 H_0
\end{equation}
where $H_0 = 100\; h$ km/s/Mpc is the Hubble constant today. On the other hand,
for the density in the  $\varphi$ field to have the correct order of magnitude
to explain the observed data we should have:
\begin{equation}
M^4 \simeq \rho_c = \frac{3 H_0^2 M_{Pl}^2}{8 \pi}.
\end{equation}
Combining these two requirements results in:
\begin{equation}
f > \frac{M_{Pl}}{\sqrt{24 \pi}}; \;\;\;\;
M \simeq 3 \times 10^{-3} \; h^{1/2} \; \mbox{eV},
\end{equation}
where we use $M_{Pl} = 1.2 \times 10^{19}$ GeV.
Such a high energy scale for the axion decay
constant is also required in order to suppress isocurvature fluctuations to
acceptable levels \cite{Dine&Anisimov,Kolb&Turner} in models where the axion is 
the dark matter. In our case, however, the axion field will only become 
dynamical in the future, behaving today as a cosmological constant and hence
there are no bounds from isocurvature fluctuations.

Recently, a two-axion model for quintessence was proposed that satisfies the 
above constraints 
in a natural fashion \cite{quintaxion}. It is based on a model independent 
superstring axion ($a_{MI}$) \cite{witten}, with a typical decay constant 
$f_{a_{MI}} = M_{Pl}$, which mixes with another axion ($a_h$) coming from  
the dynamics of the hidden sector, responsible for supersymmetry breaking. 
The hidden sector axion, with a typical decay constant given by the hidden 
sector scale, $f_{a_{h}} = 10^{13}$ GeV, arises from squark condensation 
that spontaneously breaks a global $U(1)$ symmetry, leaving supersymmetry 
intact. Non-perturbative quantum gravity effects, such as wormholes, may break
this symmetry but such effects, even if they exist, can be made exponentially
small \cite{GravityGlobal}. 

A linear combination of the two axion fields that couples
only to $SU(3)$ gauge fields is identified with the QCD axion, which solves the
strong CP problem, while its orthogonal combination is the axion responsible for
quintessence, which has a decay constant close to $M_{Pl}$. It is this
quintessential axion that we will identify with the imaginary part of a complex
scalar field  $\Phi$. The breaking of the global $U(1)$ symmetry will be
parametrized by an effective Landau-Ginzburg theory for $\Phi$.
\footnote{It should be mentioned that an effective theory description is
presumably inappropriate for such a large spontaneous symmetry breaking scale
$f$ of the order of the Plack mass, as required by the flatness of the
quintessence potential. However, it is possible to build models with extra axion
fields where this flatness can be achieved for a particular linear combination 
of the axion fields with a much smaller effective value of $f$ \cite{twoaxion}.}

In this model, supersymmetry in our world is broken by gravity mediated 
hidden sector
gaugino condensation and the height of the instanton induced potential, $M$, 
depends on the gaugino mass of TeV scale, as well as on the mass of 
hidden sector quarks. The right order of magnitude for $M$ is obtained for
parameters which solve the so called $\mu$-problem as well \cite{quintaxion}.
The squark condensate in the hidden sector would be described by
an effective Landau-Ginzburg model and $\tilde{Q} \tilde{Q}^c$ would be
identified with the complex scalar field $\Phi$. 

Another version of the two-axion model was recently proposed as a
realistic candidate for natural inflation\cite{twoaxion}, where one of the 
axions is identified with the inflaton field. 
Also, a natural candidate for the inflaton appears in models of one 
extra dimension compactified on a circle of radius R, 
where the Wilson loop for the 5th component of an abelian 
gauge field
behaves as an axion for energies below the compactification scale and its
effective decay constant can be made larger than the Planck scale for
sufficiently small 4d gauge coupling\cite{extra}. This axion model from extra
dimensions can also be used for quintessence \cite{gauge}, but at this point is
unclear if a model with a scalar complex field like the one suggested here can
be incorporated in this context.

The idea that the real part of a complex field of an Peccei-Quinn like axial 
$U(1)_{PQ}$  symmetry may drive inflation is also not new. 
Actually, Pi made this
proposal in the context of a $SU(5) \otimes U(1)_{PQ}$ model containing a complex
$SU(5)$ singlet field \cite{Pi}. There are other grand unified models, such as
$SO(10)$, where the
axion appears naturally \cite{babu&barr}. However, our model is different in the sense
that it does
not start with $\langle \Phi \rangle = 0$, but rather has an initial value 
such that  $\langle \Phi \rangle \gg f$, as in a chaotic inflation model
\cite{Linde_chaotic}.

In fact, at large values of the field, $|\Phi| \gg f$, the model is basically a 
$\lambda|\Phi|^4$ model. The slow roll conditions are satisfied for 
$|\Phi| > M_{Pl}/\sqrt{2 \pi}$ and
the Universe undergoes a phase of chaotic inflation, in which it expands and
cools down rapidly. 
In order to have a sufficient number of e-foldings, the
initial value of the field must be $|\Phi|_i \gtrsim 5 M_{Pl}$.
The self-coupling
constant must satisfy $\lambda \lesssim 10^{-15}$ to generate the correct
amplitude
of density perturbations.\footnote{One may rightly wonder whether the potential 
can be
protected against quantum gravity corrections for this large field value. It has
been shown that introducing the shift symmetry present in axion models allows a
chaotic inflation model to be embedded in a supergravity model with naturally
large field values\cite{supergravity}.}
One should keep in mind that, quoting Kinney \cite{Kinney},
it is fair to say that $\lambda \phi^4$ inflation is under
significant pressure from the data but it cannot be definitively ruled out.

At this stage we can use the classical equations of motion to 
study the dynamics:
\begin{equation} \label{eq:eta}
\ddot{\eta} + 3 H \dot{\eta} - \frac{\dot{\varphi}^2}{f^2} \eta + V'(\eta) = 0
\end{equation}
and
\begin{equation} \label{eq:phi}
\ddot{\varphi} + \left(3 H + \frac{\dot{g}}{g} \right) \dot{\varphi} +
\frac{1}{g(\eta)} V'(\varphi) = 0
\end{equation}
where $g(\eta) = \eta^2/f^2$ and
\begin{equation}
V(\eta) = \frac{\lambda}{4} (\eta^2 - f^2)^2 ; \;\;\; 
V(\varphi) =  M^4 [ \cos(\varphi/f) -1 ]
\end{equation}
Neglecting the instanton potential it follows that
\begin{equation} 
\dot{\varphi} \propto \frac{1}{a^3 g(\eta)} 
\end{equation}
and therefore the field $\varphi$ is frozen during inflation.

As the field $\eta$ rolls down the potential, the slow-roll conditions 
eventually cease to be
valid, inflation ends and the Universe is
effectively at zero temperature. 
For $|\Phi| \simeq f$, the systems becomes sensitive
to the symmetry breaking process. 
The dynamics of spontaneous symmetry breaking (SSB) in the early Universe is
complicated by the presence of large quantum fluctuations due to the tachyonic
instability and it has been
studied in detail by Felder {\it et al.} \cite{SSB}. 

The difference between our model and the one studied by Felder et al. lies in the fact that in their case the field rolls from the central hill of the ``mexican hat" potential down to its valley while in our case the field rolls from the border of the hat towards its valley. Hence, in our case the tachyonic instability does not appear
unless a large amount of fine tuning is invoked in our initial conditions to put the field exactly at the top of the center of the potential, which is not natural.
Therefore, what follows is really a preheating akin to chaotic models with quadratic potentials. 
  
Energy can be efficiently transferred from the field to
other particles, resulting in the reheating of the
Universe. 
Reheating can occur via couplings like $g \Phi \Phi^\ast h h $ or 
$\frac{g^\prime}{M} \Phi \Phi^\ast \bar{\psi} \psi$, where $h$ is a scalar field and
$\psi$ is a fermionic field. The coupling to fermions must be a dimension-5
operator due to the $U(1)$ symmetry and hence is suppressed by a high energy
scale $M$. These couplings can introduce resonances which make dominant the first oscillation of the field, as it happens in no-oscillation models \cite{us}.

A detailed analysis of reheating requires a more definite model regarding interactions with Standard Model fields and is beyond the scope of this paper.
However, we mention that reheating due to coupling to massless scalars in a similar model has been studied in \cite{us} and results in a reheating temperature of the order of $T_{rh} = 0.1 (\lambda^{3/4} g^{15/2})^{1/4} \; M_{Pl}$. In order to avoid large radiative
corrections to the potential, which are of the order of $g^2/16 \pi^2$, the
coupling constant $g$ must be smaller than $g \sim 10^{-6}$. Using $\lambda =
10^{-14}$ and  $g = 10^{-6}$ one obtains a reheating temperature $T_{rh} = 
10^4$ GeV, avoiding the gravitino problem.

A numerical study based on 
non-perturbative lattice
simulations shows that the field is quickly trapped at the minimum of the 
potential,  $|\Phi| = f/\sqrt{2}$. The physical heavy scalar $\eta$ gets a mass of
$m_{\eta} = \sqrt{2 \lambda} f \simeq 5\times 10^{11}$ GeV, too heavy to be
produced in any foreseeable accelerator.
The global axionic strings Felder {\it et al.} find in their simulation will not 
appear in our
model since there is inflation before SSB and the phase of the field, which 
is frozen at some random value, will be uniform throughout the observable 
Universe.

In our model, the field $\varphi$ will become dynamical at a time $t_\ast$ 
when $m_{\varphi} \simeq 3 H(t_\ast)$. When this happens, the dark energy 
will be transferred to 
axion production arising from the field coherent oscillation down the 
instanton induced potential. Therefore, cosmic acceleration will stop in the
future, avoiding the problem with event horizons in string theory
\cite{strings}. 

Finally, we would like to mention that one may be concerned that quantum fluctuations could generate large perturbations in the angular 
$\varphi$ field. 
We showed that the classical 
zero-momentum mode of the angular field is frozen at early times 
(that is, until well after reheating); as a result, any initial spatial 
gradients in $\varphi$ will be smoothed out by inflation, providing nearly 
homogeneous initial conditions for late-time acceleration. Quantum effects 
can, however, lead to inhomogeneities in $\varphi$ in two ways: through quantum 
fluctuations generated during inflation or via coupling to the 
radial $\eta$ field during reheating. 

It is well known that quantum fluctuations in $\varphi$ during inflation can generate isocurvature fluctuations  \cite{DT}. In fact, dark energy isocurvature fluctuations were recently suggested as a mechanism to reduce the power spectrum seen at the quadrupole scale in the cosmic microwave background \cite{GH}.
A model similar to ours but with tachyonic preheating was invoked to amplify the dark energy perturbations and at the same time reducing a large amount of gravitational wave signal that could potentially invalidate the mechanism \cite{GW}.
However, in our model these quantum fluctuations are not a cause of concern, 
since they are of order $H_{inf}$, the Hubble 
parameter during inflation, which is much smaller than the classical 
field amplitude, of order $f$, by a factor of order $\sqrt{\lambda}$; these 
would give rise to isocurvature perturbations in $\varphi$, but only at very 
late times, and are too small to affect the homogeneous dynamics of $\varphi$ 
during the late acceleration phase. 

As for inhomogeneities generated at reheating, eqns. (7) and (8) 
show that the coupling between $\varphi$ and $\eta$ 
is strongly suppressed by inverse powers of $f$, which is of order the 
Planck scale. In fact, the interaction cross section of 
$\varphi$ with other fields is generally of order $1/f^2$, so it 
decouples at a temperature of order $f$. In any event, such interactions 
will only produce very small-scale fluctuations of $\varphi$, on scales of order 
the Hubble radius or smaller during reheating. For late-time acceleration, 
we are concerned with the variations of $\varphi$ on the scale of the present 
Hubble radius, which is many orders of magnitude larger than the 
comoving Hubble radius during reheating. A detailed study of reheating in 
this model is beyond the scope of this article 
and will be the subject of a future paper. Nevertheless, the arguments 
above indicate that these effects will not lead to large-scale 
inhomogeneities in the angular field that would impede its driving 
late acceleration. 

\section{Conclusion}

In conclusion, we have presented in this article a simple, well-motivated model for 
quintessential inflation with a complex field. This model unifies the successes
of a chaotic inflation model and a pseudo-goldstone boson quintessence model into a 
single framework and naturally relates both phases of cosmic acceleration.

The model suggested here is certainly an incomplete, exploratory toy model. 
The coupling $\lambda$ would be naturally small, in the sense of
$\,$  't Hooft, if the model is embedded in supergravity-like models, 
as shown in \cite{supergravity}. 
The smallness of the instanton scale $M$ is related to parameters in the hidden
sector in models of gravity mediated supersymmetry breaking.
It would be highly desirable to build a realistic model based on these ideas.

\section*{Acknowledgments}
RR would like to thank the hospitality at the Fermilab Theory Group, where this 
work was initiated when he was a Summer Visitor. JAF thanks C. T. Hill and H. P. Nilles for
discussions. RR thanks A. Linde and C. Gordon for providing interesting
comments. The work of RR was supported by grants from Fapesp and CNPq 
(brazilian agencies) and JAF acknowledges support from the
NSF Center for Cosmological Physics at the University of Chicago, the DOE at
Fermilab and Chicago, and NASA grant NAG 5-7092 at Fermilab.
%
%

\end{document}